\documentclass{aa}         
\usepackage{txfonts}
\usepackage{psfig}
\usepackage{natbib}
\bibpunct{(}{)}{;}{a}{}{,}%

\begin{document}
\titlerunning{Diagnostics of stellar flares}
\authorrunning{Fabio Reale\inst{1,2}}

\title{Diagnostics of stellar flares from X-ray observations: from the decay to the rise phase}
\author{Fabio Reale\inst{1,2}}
\institute{Dipartimento di Scienze Fisiche \& Astronomiche, Sezione di
Astronomia, Universit\`a di Palermo, Piazza del Parlamento 1,
I-90134 Palermo, Italy
\email{reale@astropa.unipa.it}
\and
INAF - Osservatorio Astronomico di Palermo ``Giuseppe
S. Vaiana", Piazza del Parlamento 1, I-90134 Palermo, Italy
}

\abstract{The diagnostics of stellar flaring coronal
loops have been so far largely based on 
the analysis of the decay phase. }
{We derive new diagnostics from the analysis of the
rise and peak phase of stellar flares.}
{We release
the assumption of full equilibrium of the flaring loop at the flare peak,
according to the frequently observed
delay between the temperature and the density maximum.
From scaling laws and hydrodynamic simulations we derive
diagnostic formulas as a function of observable quantities and times.}
{We obtain a diagnostic toolset related to the rise phase, 
including the loop length, density and aspect ratio.
We discuss the limitations of this approach
and find that the assumption of loop equilibrium in the
analysis of the decay leads to a moderate
overestimate of the loop length. A few relevant applications
to previously analyzed stellar flares are shown. }
{The analysis of the flare rise and peak phase 
complements and completes the analysis of the decay phase.}

\maketitle

\section{Introduction}

Solar and stellar coronal flares are impulsive events well detected in
the soft X-ray band. They are typically explained as due to the sudden
increase of temperature and emission measure of the plasma confined in
single or groups of magnetic tubes (loops), caused by strong 
heat pulses.

The flare light curves in the soft X-rays are typically characterized by
a fast rise phase followed by a slower decay (e.g. Haisch et al. 1983).
In the decay phase, the plasma cooling is due to radiation emission and
to thermal conduction to the chromosphere. Since the cooling times for a
confined plasma depend on the characteristics of the confining structure,
and in particular on its length, the analysis of the decay phase has been
extensively used to diagnose the flaring loops, and in particular their
size (see Reale 2002 for an extensive review). This is important not only
because stellar flaring regions are unresolved, but also because, more
in general, this is one of the few tools to obtain detailed information
on the geometry of the stellar coronae (e.g. Favata et al. 2000a), and
of any other phenomena involving plasma confined in magnetic structures.

To summarize, Serio et al. (1991) derived a thermodynamic time scale
for the pure cooling of flaring plasma confined in single coronal loops:

\begin{equation}
\tau_{s} = 3.7 \times 10^{-4} \frac{L}{\sqrt{T_0}}
= 120 \frac{L_9}{\sqrt{T_{0,7}}}
\label{eq:tserio}
\end{equation}
where $L$ ($L_9$) the loop half-length (in units of $10^9$
cm), $T_0$ ($T_{0,7}$) the loop maximum temperature (in units of $10^7$
K). In principle, the loop length can be derived simply by inverting
Eq.(\ref{eq:tserio}).  Jakimiec et al. (1992) and Sylwester et al. (1993)
analyzed extensively the decay of solar coronal flares and pointed out
that significant heating can be released even during the late phases of a
flare. This can be diagnosed through the analysis of the path of the decay
in a density-temperature diagram: sustained heating slows down the plasma
cooling but much less the density decrease. If this effect is not properly
taken into account, the loop length can be significantly overestimated.
Reale et al. (1997) use the scale time Eq.(\ref{eq:tserio}) to derive a
formula for loop length, corrected to include the effect of significant
heating in the decay (see the Appendix for a review).  This approach has
been extensively applied to analyze observed stellar flares (see Reale
2003 for a review), with the exception of flares where the residual
heating completely drives the flare decay over the plasma cooling.
These flares are more appropriately described with models of two-ribbon
flares (Kopp \& Poletto 1984).

Eq.(\ref{eq:tserio}) and all the subsequent derivations lie on the
assumption that the flare decay starts when the loop is at hydrostatic and
energy equilibrium.  Extensive modeling of solar and stellar flares has
shown that the heat pulses that trigger the flare are impulsive, i.e.
their duration is small with respect to the overall flare duration.
One then wonders whether the flaring loop has had enough time to reach
equilibrium before the heat pulse is switched off or not,
and, if not, whether this might be important for the analysis of the
decay.  Jakimiec et al. (1992) showed that temperature and density begin
to decrease simultaneously if the heating lasts long enough to reach
equilibrium.  In many flares, it is instead observed that the temperature
peaks (and therefore begins to decrease) measurably before the emission
measure, both in solar flares (e.g. Sylwester et al. 1993) and in stellar
flares (van den Oord et al. 1988, van den Oord et al. 1989, Favata et
al. 1999, Favata et al. 2000b, Stelzer et al. 2002).  Cargill \& Klimchuk
(2004) pointed out that in transiently heated loops the cooling is
initially dominated by thermal conduction and that the density begins to
decay as soon as the radiative and conduction cooling times become equal.

Most attention has been so far dedicated to the decay phase,
which is the longest-lasting part of flares and therefore typically
offers more opportunities of time-resolved data analysis and of good
photon statistics. Little and partial attention has been instead devoted
to the initial phase of the flares.  Hawley et al. (1995) study the rise
phase jointly to the decay phase, including neither the heating in the
decay, nor the delay between the temperature and the density peak.

Flare observations from recent missions such as Chandra and XMM-Newton
often detect flares in great detail and succeed in resolving the rise
phase (e.g. G\"{u}del et al. 2004). 
On the other hand, in long flares it happens that the observations
ends early in the decay phase, inhibiting the related analysis and
making any kind of information derivable from the rise phase important (e.g.
Giardino et al. 2004).

In this work, we address specifically the rise and peak phase of flares,
and investigate what diagnostics can be extracted from its analysis.
We will study the flare initial phases as a stand-alone analysis, and
compare and cross-check with the analysis of the decay. We will also
address possible additional diagnostics, e.g.  the density and the loop
aspect ratio cross-section radius over length -- are flaring loops fat
or slim or arcades of loops? -- which cannot be constrained just from the
decay analysis.  In our derivation, we will maintain the assumption that
the flare occurs in a single loop. This assumption is more realistic
in the rise phase, when the impulsive heating typically involves a
dominant loop structure while later residual heating may be released
in other similar adjacent loops (e.g. Aschwanden \& Alexander 2001).
We will instead release the assumption that the flaring loop evolves to
a condition of equilibrium, and therefore also address the
question of what is the effect of releasing equilibrium conditions on
diagnostic formulae, i.e. if the decay starts before the loop reaches
equilibrium conditions, and what is the error from assuming equilibrium
conditions. We will derive the loop length from the rise phase.

In Section~\ref{sec:anal}, we analise the flare evolution
with a general outline, operative definitions, relevant results from
modeling. In Section~\ref{sec:deriv}, diagnostic tools
for the analysis of the flare rise and peak phase are derived. In Section~\ref{sec:disc},
we discuss the results, the limitations of the analysis, the applications,
with some specific examples and in Section~\ref{sec:conc} we draw
our conclusions.

\section{Flare analysis}
\label{sec:anal}

\subsection{General flare evolution}
\label{sec:fl_ev}

We consider a flare occurring in a single coronal loop. A flaring coronal loop
can be modelled as a magnetic flux tube where the plasma is heated to
flare temperatures by a transient heat pulse. The plasma confined in the
loop can be described as a compressible fluid which moves and transports
energy along the magnetic field lines (e.g. Priest 1984).

We will suppose that: (i) the flare occurs in a semicircular loop with
half-length $L$, initially in equilibrium conditions at much lower
temperature and density than at flaring conditions; (ii) the flare is
triggered by a heat pulse uniformly distributed in the loop; (iii) the
heat pulse is a top-hat function in time; (iv) there is no heating in
the decay; (v) the flaring loop is much smaller than the pressure scale
height at the flare temperature.

\begin{figure}
\centerline{\psfig{figure=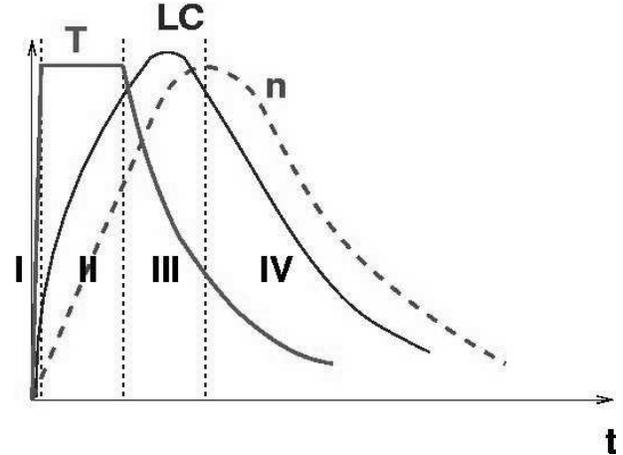,width=8cm}}
\caption[]{Scheme of the evolution of temperature (T, {\it thick solid line}), X-ray emission (LC, {\it thinner solid line}) and density (n, {\it dashed line})
during a flare triggered in a coronal loop by a 
heat pulse. The flare evolution is divided into four phases (I, II, III, IV, see text for further details).
\label{fig:ntx_time}}
\end{figure}

The plasma cooling is governed by the thermal conduction to the
cool chromosphere and by radiation from optically thin conditions.
The evolution of the confined plasma is well-known from observations
and from modeling (e.g. Nagai 1980, Peres et al. 1982, Cheng et al.
1983, Nagai \& Emslie 1984, Fisher et al.  1985, MacNeice 1986, Betta
et al. 2001) and in the following we summarize it into four phases,
sketched in Fig.~\ref{fig:ntx_time}, which map on the path drawn in
the density-temperature diagram of Fig.~\ref{fig:nt_diag_eq} (see also
Jakimiec et al. 1992).

\begin{description}

\item[Phase I]: from the start of the heat pulse to the temperature peak
({\it heating}).  The heat pulse is triggered in the loop and the heat is
efficiently conducted down to the much cooler and denser chromosphere.
The temperature rapidly increases in the whole loop, with a time scale
given by the conduction time in a low density plasma (see below).

\item[Phase II]: from the temperature peak to the end of the heat
pulse ({\it evaporation}). The temperature settles to the maximum value ($T_0$).
The chromospheric plasma is strongly
heated and expands upwards, filling the loop with much denser plasma.
The evaporation is explosive at first, with a timescale given by the
isothermal sound crossing time:

\begin{equation}
\tau_{sd} = \frac{L} {\sqrt{2 k_B T_0 / m}} \approx 25 \frac{L_9}{ \sqrt{T_{0,7}}}
\label{eq:tsound}
\end{equation}
where $k_B$ is the Boltzmann constant, $m$ is the average particle mass.
After the evaporation front has reached the loop apex, the loop 
continues to fill more gently. The time scale during this
more gradual evaporation is dictated by the time taken by the cooling
rate to balance the heat input rate (see Sec.~\ref{sec:deriv}).

\item[Phase III]: from the end of the heat pulse to the density peak
({\it conductive cooling}).  When the heat pulse stops, the plasma
immediately starts to cool due to the efficient thermal conduction
(e.g. Cargill \& Klimchuk 2004), with a time scale:

\begin{equation}
\tau_c=\frac{3 n_c k_B T_0 L^2}{2/7\kappa T_0^{7/2}}
= \frac{10.5 n_c k_B L^2}{\kappa T_0^{5/2}} \approx 50 \frac{n_{c,10} L_9^2}{T_{0,7}^{5/2}}
\label{eq:tcon}
\end{equation}
where $n_c$ ($n_{c,10}$) is the particle density ($10^{10}$ cm$^{-3}$) at
the end of the heat pulse,
$\kappa = 9 \times 10^{-7}$ (c.g.s. units) is the thermal conductivity.

The heat stop time can then be generally traced as the time at which the
temperature begins to decrease significantly and monotonically. While
the conduction cooling dominates, the plasma evaporation is still going
on and the density increasing.  The efficiency of radiation cooling
increases as well. On the other hand, the efficiency of conduction
cooling decreases with the temperature.

\item[Phase IV]: from the density peak afterwards ({\it Radiative
cooling}).  As soon as the radiation cooling time becomes equal to the
conduction cooling time (Cargill \& Klimchuk 2004), the density reaches
its maximum, and the loop depletion starts, slow at first and then
progressively increasing. The pressure begins to decrease inside the loop,
and is no longer able to sustain the plasma. In this phase, radiation
becomes the dominant cooling mechanism, with the following time scale:

\begin{equation}
\tau_r = \frac{3 k_B T_M}{n_M P(T)} =  \frac{3 k_B T_M}{b T_M^\alpha n_M} \approx
9 \times 10^{3} \frac{T_{M,7}^{3/2}} {n_{M,10}}
\label{eq:trad}
\end{equation}
where $T_M$ ($T_{M,7}$) is the temperature at the time of the density
maximum 
($10^7$ K), $n_M$ ($n_{M,10}$) the maximum density ($10^{10}$ cm$^{-3}$),
$P(T)$ the plasma emissivity per unit emission measure, expressed as:
\[
P(T) =  b T^\alpha
\]
with $b = 1.5 \times 10^{-19}$ and $\alpha = -1/2$,
for consistency with the parameters of the loop scaling laws (Rosner
et al. 1978). In this phase, the density and the temperature
both decrease monotonically. The presence of significant residual heating
could make the decay slower. This can be diagnosed from the analysis of the
slope of the decay path in the density-temperature diagram 
(Sylwester et al. 1993, Reale et al. 1997, see the Appendix).

\end{description}

\subsection{Heat pulse duration}

\begin{figure}
\centerline{\psfig{figure=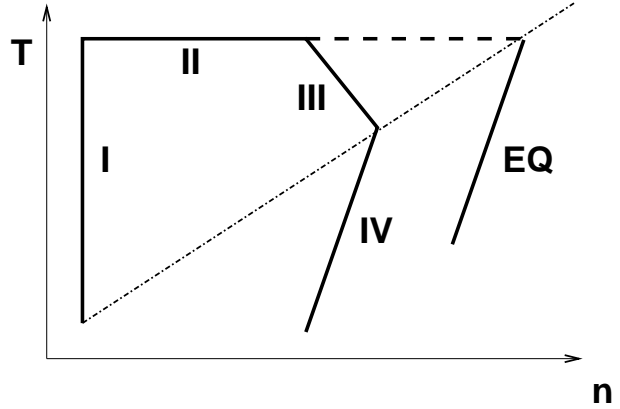,width=8cm}}
\caption[]{Scheme of the flare evolution of Fig.~\ref{fig:ntx_time} in
a density-temperature diagram ({\it solid line}). The four phases are
labelled. The locus of the equilibrium loops is shown ({\it dashed-dotted
line}), as well as the flare path with an extremely long heat pulse
({\it dashed line}). The corresponding decay path (marked with EQ) is
the one typically considered by flare analysis based on the decay phase.
\label{fig:nt_diag_eq}}
\end{figure}

The evolution outlined in Sec.~\ref{sec:fl_ev} concerns a flare driven
by a transient heat pulse. The analysis of stellar flares based on the
decay phase typically assumes that the decay starts from equilibrium
conditions, i.e. departing from the the locus of the equilibrium
loops with a given length (hereafter QSS line, Jakimiec et al. 1992)
in Fig.~\ref{fig:nt_diag_eq}. The link between this assumption and the
flare evolution outlined above is shown in Fig.~\ref{fig:nt_diag_eq}:
if the heat pulse lasts long enough, phase II extends to the right,
and the flaring loop asymptotically reaches equilibrium conditions, i.e.
the horizontal line approaches the QSS line. It is worth noting that,
if the decay starts from equilibrium conditions, Phase III is no longer
present, and Phase II links directly to Phase IV. Therefore, there is no
delay between the beginning of the temperature decay and the beginning
of the density decay: the temperature and the density start to decrease
simultaneously. Also, the decay will be dominated by radiation cooling,
except for its very beginning (Serio et al. 1991).

The analysis of the rise and peak phase has to include the presence of
Phase III, and the delay between the temperature peak and the density
peak. This delay is often observed both in solar flares (e.g. Sylwester
et al. 1993) and in stellar flares (e.g. van den Oord et al. 1988,
1989, Favata et al. 2000b, Maggio et al. 2000, Stelzer et al. 2002). The
presence of this delay is a signature of a relatively short heat pulse,
or, in other words, of a decay starting from non-equilibrium conditions.

\subsection{Hydrodynamic modeling}
\label{sec:hyd}

We now use hydrodynamic simulations to analyze more
in detail the evolution of the rise and peak phase of a flare (Phases I
to III).  The Palermo-Harvard code solves the time-dependent hydrodynamic
equations for the plasma confined in a loop to describe the evolution
of the density, temperature and velocity of the plasma along the loop
(Peres et al. 1982, Betta et al. 1997).

As a representative example, we consider an initially quiet coronal
loop with half-length $L = 2 \times 10^9$ cm and temperature of about 3
MK. A transient heat pulse is injected in it at time t=0.  The flare heat
pulse lasts for a time $t_{heat}$, is uniformly distributed in the loop,
and is as intense (9 erg cm$^{-3}$ s$^{-1}$)
as to bring the loop to a temperature of $\sim 20$ MK.

\begin{figure}
\centerline{\psfig{figure=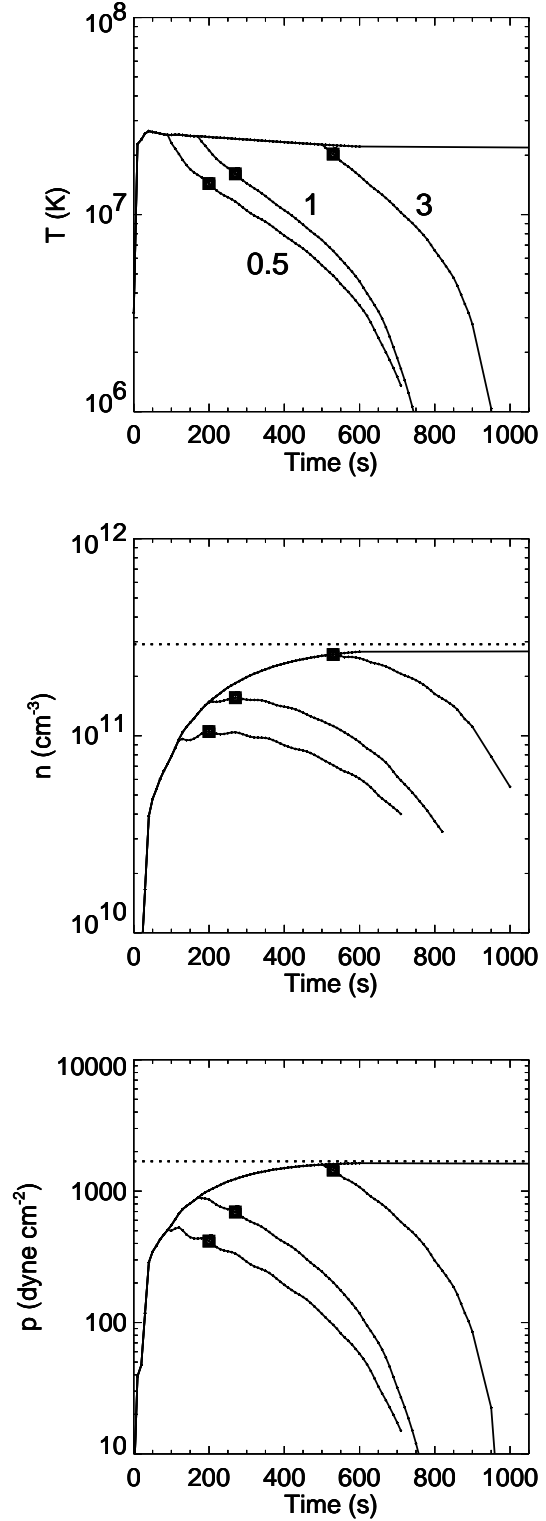,width=8cm}}
\caption[]{Evolution of the plasma temperature ({\it top panel}),
density  ({\it middle}) and pressure ({\it bottom}) at the loop apex,
as computed from a hydrodynamic model of a 20 MK flare in a loop with
half-length $L = 2 \times 10^9$ cm, and with three different durations
of the heat pulse (labelled in the top panel, in units of $\tau_s$) and with non-stopping
heating. The asymptotic density and pressure values are also shown ({\it
dotted horizontal lines}).  The time of the density maximum is marked
({\it black spots}).
\label{fig:mod_evol}}
\end{figure}

Fig.~\ref{fig:mod_evol} shows the evolution of the plasma temperature,
density and pressure at the loop apex for three different durations of
the heat pulse (i.e. $t_{heat}=90$ s $\approx 0.5 \tau_s$, $t_{heat}=170$
s $\approx \tau_s$ and $t_{heat}= 500$ s $\approx 3 \tau_s$) and for
non-stopping heating. As long as the heat pulse is on, the simulation
results overlap; they differ only for the decay phase, when the
heating is off.  The decay of the simulations with longer-lasting heat
pulses begins later. As already mentioned in Sec.~\ref{sec:fl_ev},
the temperature begins to decrease as soon as the heat pulse stops.
The density and the pressure, instead, both continue to increase and they
reach their maximum well later.  We note that the shorter the heat pulse,
the longer is the delay between the beginning of the temperature decay
and the density maximum.  For the longest-lasting heat pulse, the density
and pressure values are very close to the asymptotic equilibrium values,
estimated from loop scaling laws (Rosner et al. 1978).
We have checked that the evolution is self-similar for a longer loop $L
= 10^{10}$ cm at the same temperature, with the evolution times scaling
as $\tau_s$.

\begin{figure}
\centerline{\psfig{figure=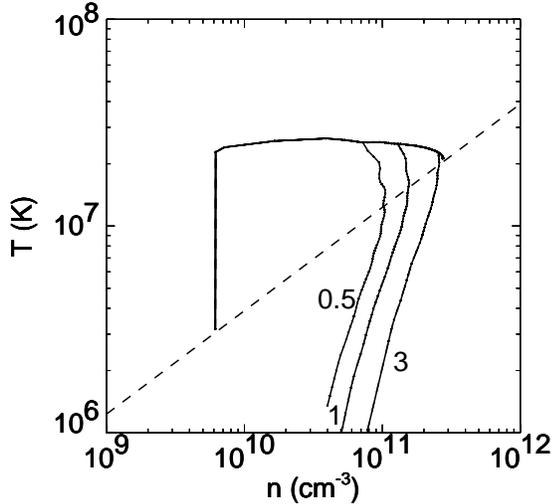,width=8cm}}
\caption[]{Evolution of the flare of Fig.~\ref{fig:mod_evol} in the density-temperature
diagram. The locus of the equilibrium loops derived from loop
scaling laws (QSS curve, {\it dashed line}) is also shown.
\label{fig:mod_nt}}
\end{figure}

In the density-temperature diagram (Fig.~\ref{fig:mod_nt}), the flare
evolution for the different heat pulse durations is well in agreement
with that sketched in Fig.~\ref{fig:nt_diag_eq}. For short-lasting pulses,
phase IV starts as soon as the path crosses the QSS curve.  For $t_{heat}
\approx 3 \tau_s$, phase II ends and the decay (phase IV) starts both very
close to the QSS curve, while phase III is practically absent (Jakimiec
et al. 1992).

\section{Diagnostics of the rise and peak phase}
\label{sec:deriv}

\begin{figure}
\centerline{\psfig{figure=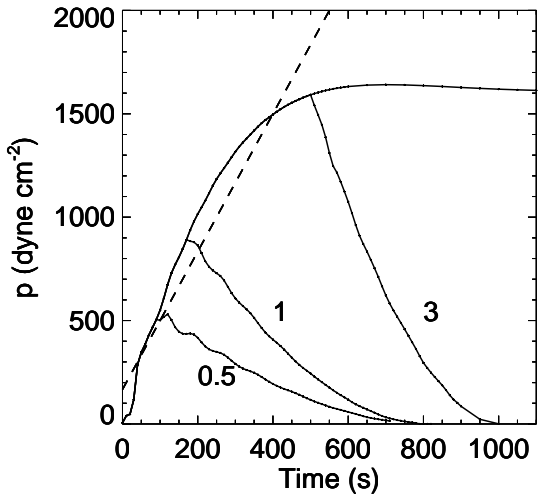,width=8cm}}
\caption[]{Pressure evolution of the flare of Fig.~\ref{fig:mod_evol} in a linear scale 
to show that most of the rise phase can be reasonably described with a
linear trend ({\it dashed lines}).
\label{fig:mod_lin}}
\end{figure}

We now derive simple diagnostic tools for the analysis of the rise and
decay phase of a flare, taking advantage of detailed numerical modeling.
We first observe that the maximum possible duration of the rise phase
is the time taken by the loop to reach equilibrium conditions under
the action of a constant (flare) heating.  The simulation results
in Fig.~\ref{fig:mod_evol} show that the time taken by the plasma to
reach equilibrium conditions is much longer than the sound crossing time
(Eq.[\ref{eq:tsound}]), which rules the very initial plasma evaporation.
This is also clear in Fig.~\ref{fig:mod_lin}, which shows the evolution
of the pressure at the loop apex in a linear scale: after $t = 50$ s,
the rate of pressure enhancement becomes more gentle.  As mentioned
in Sec.~\ref{sec:fl_ev}, in this phase the dynamics become much less
important and the interplay between cooling and heating processes
becomes dominant. The relevant time scale is therefore that reported
in Eq.~(\ref{eq:tserio}).

After detailed analysis of extensive numerical modeling of
decaying flaring loops, we have checked that the decay time from
Eq.~(\ref{eq:tserio}) should be computed with a correction factor:
\begin{equation}
\tau^\prime_s = \phi \tau_s
\label{eq:taunew}
\end{equation}
where $\phi \approx 1.3$ to better fit the decay time measured from
numerical models.

\begin{figure}
\centerline{\psfig{figure=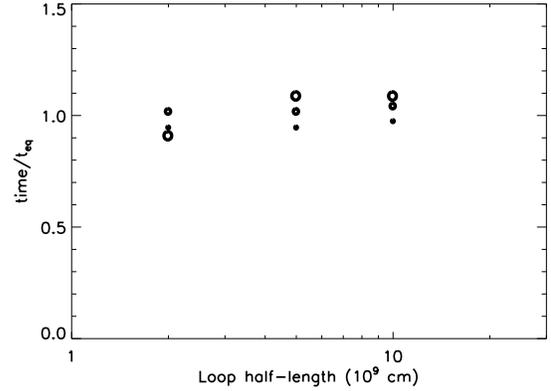,width=8cm}}
\caption[]{Time to reach pressure equilibrium obtained from hydrodynamic
simulations: the ratio of the time at which the pressure is  97\% of the
equilibrium value to the time obtained from Eq.~(\ref{eq:tequil}) vs loop
length. The size of the data points is proportional to the heating rate
(maximum temperature of 10, 20 and 30 MK).
\label{fig:tequil}}
\end{figure}

Hydrodynamic simulations confirm that
the time required to reach full equilibrium scales as the loop cooling time ($\tau_s$), and, as shown for instance in
Fig.~\ref{fig:mod_lin} (see also Jakimiec et al. 1992),
the time to reach flare steady-state equilibrium is:

\begin{equation}
t_{eq} \approx 3 \tau_s \approx 2.3 \tau^\prime_s
\label{eq:tequil}
\end{equation}
We have verified that Eq.(\ref{eq:tequil}) holds for loops of different
lengths. Fig.~\ref{fig:tequil} shows the ratio of the time required to
reach 97\% of the pressure equilibrium value to the time computed
from Eq.~(\ref{eq:tequil}) for simulations of flaring loops with three
different loop lengths, and three different heating rates appropriate to
reach the temperature of 10, 20 and 30 MK, respectively. The agreement 
is within 10\%.

For $t \geq t_{eq}$, the density asymptotically approaches the equilibrium
value:

\begin{equation}
n_{0} = \frac{T_0^2}{2a^3 k_B L } = 1.3 \times 10^6 ~ \frac{T_0^2} {L}
\label{eq:n0}
\end{equation}

\noindent
where $a = 1.4 \times 10^3$ (c.g.s. units), or

\[
n_{0,10} = 13 \frac{T_{0,7}^2}{ L_9}
\]
as obtained from the loop scaling laws (Rosner et al. 1978) and the 
plasma equation of state.

If the heat pulse stops before the loop reaches equilibrium conditions,
the flare maximum density is lower than the value at equilibrium.
Fig.~\ref{fig:mod_lin} shows that, after the initial impulsive evaporation
on a time scale given by Eq.(\ref{eq:tsound}), the later progressive
pressure growth can be approximated with a linear trend. Since the
temperature is almost constant in this phase (Fig.~\ref{fig:mod_evol}),
we can as well approximate that the density increases linearly for most
of the time. We can then estimate the value of the maximum density at
the loop apex as:

\begin{equation}
n_{M} \approx n_0 \frac{t_{M}}{t_{eq}}
\label{eq:nmax}
\end{equation}
where $t_{M}$ is the time at which the density maximum occurs, which
can be measured directly from observations.  Since it is reasonable
to assume that the volume of the flaring region does not change much,
at least on the relatively short time scale of the rise phase, the time
of the maximum emission measure is a good proxy for $t_M$.

Phase~III ranges between the time at which the heat pulse ends and
the time of the density maximum. Fig.~\ref{fig:mod_evol} shows that
the start time of the temperature decay marks well the end of the
heat pulse. In many stellar flares, the temperature evolution is not
well resolved, and we may take the time of the temperature maximum as
indicative of the end of the heat pulse.

The time of the density maximum is also the time at which the decay path
crosses the locus of the equilibrium loops (QSS curve).  As reported
in Sec.~\ref{sec:fl_ev}, Cargill \& Klimchuk (2004) remarked that, at
this time, the radiative cooling time is exactly equal to the conductive
cooling time.  By equating the time scales in Eqs.~(\ref{eq:trad}) and
(\ref{eq:tcon}), we can then derive the temperature $T_{M}$ at which
the flare maximum density occurs:

\begin{equation}
T_{M} = 9 \times 10^{-4} ( n_M L )^{1/2}
\label{eq:trc}
\end{equation}
or
\[
T_{M,7} = 0.28 ( n_{M,10} L_9 )^{1/2}
\]
If a value for $L$ is already available to us and we are able to measure
$T_M$, we can derive $n_M$ from Eq.(\ref{eq:trc}), in alternative to
Eq.(\ref{eq:nmax}).

Since phase III is dominated by conductive cooling, we derive its
duration, i.e. the time from the end of the heat pulse to the density
maximum, as

\begin{equation}
\Delta t_{0-M} \approx \tau_c \ln \psi 
\label{eq:dt0rc}
\end{equation}
where
\[
\psi = \frac{T_0}{T_{M}} 
\]
and $\tau_c$ (Eq.[\ref{eq:tcon}]) is computed for an appropriate value of
the density $n_c$.  A good consistency with numerical simulations is obtained
for $n_c = (2/3) n_M$.

The collection of Eqs.~(\ref{eq:tequil}) -- (\ref{eq:dt0rc}) provides a
set of diagnostic tools for the analysis of the rise and peak phase of
stellar flares.  Eqs.~(\ref{eq:tequil}) -- (\ref{eq:nmax}) are related
to the rise phase, the others to the peak phase, or, more precisely,
phase III as defined in Sec.~\ref{sec:fl_ev}. We have checked that the
equations provide values consistent with those obtained from accurate
numerical modeling within few percent.

By combining Eq.(\ref{eq:dt0rc}) and Eq.(\ref{eq:nmax}) we obtain:

\begin{equation}
\frac{\Delta t_{0-M}}{t_M} \approx 1.2 \ln \psi
\label{eq:trat}
\end{equation}
which ranges between 0.2 and 0.8 for typical values of $\psi$ (1.2 -- 2).

By combining Eqs.(\ref{eq:n0}), (\ref{eq:nmax}) and (\ref{eq:trc}),
we derive a new expression for the loop half length:

\begin{equation}
L_9 \approx 3 ~ \psi^2 T_{0,7}^{1/2} t_{M,3}
\label{eq:lris}
\end{equation}
where $t_{M,3}$ is $t_M$ in units of $10^3$ s. For typical values of
$\psi$ and $T_0$, we obtain $L_9 \sim 5 - 25 t_{M,3}$ and $L_9 \sim 10
t_{M,3}$ may be taken for rough estimates. Therefore, we expect that
flares occurring in loops with length of the order of $10^{10}$ cm show
the peak of the emission measure about 1 ks after the beginning.

By including Eqs.(\ref{eq:trat}) into Eq.(\ref{eq:lris}),
we derive another alternative expression for the loop half length:

\begin{equation}
L_9 \approx 2.5 ~ \frac{\psi^2}{\ln \psi} T_{0,7}^{1/2} \Delta t_{0-M,3}
\label{eq:ldelt}
\end{equation}
where $\Delta t_{0-M,3}$ is $\Delta t_{0-M}$ in units of $10^3$ s. 
For typical values of $\psi$ and
$T_0$, we obtain $L_9 \sim 20 - 100 \Delta t_{0-M,3}$ and 
$L_9 \sim 50 \Delta t_{0-M,3}$ may be taken for rough estimates. 

The loop length derived from application of Eqs.(\ref{eq:lris}) and
(\ref{eq:ldelt}) to the model simulations is correct within $10\%$.
Eqs.(\ref{eq:lris}) and (\ref{eq:ldelt}) allow us to estimate the length
of the flaring loop from measuring characteristic time intervals of the
flare rise phase and related temperatures.

\section{Discussion: implications and applications}
\label{sec:disc}

\subsection{Consistency and limitations}

The aim of this work is to investigate the diagnostics of the rise
and peak phase of coronal flares, to complement the well-established
diagnostics of the decay phase (e.g. Reale et al. 1997).  Once derived
the relevant diagnostic expressions, we first check whether there are
effects on the analysis of the decay. In particular, we wonder whether
assuming that the decay starts from loop equilibrium conditions --
therefore ignoring the details of the ``previous history'' -- leads to
significant systematic errors or not. In the decay analysis, one 
important parameter is the temperature at the start of the decay. 
At equilibrium conditions, the decay starts at the temperature maximum.
In Sec.~\ref{sec:fl_ev}, we have pointed out that, in an impulsive
flare event, in which the loop does not reach equilibrium conditions,
the density begins to decay later than the temperature.  The proper
decay phase begins at the density peak (i.e. the later time), and one
should then use the temperature at the density peak, lower than the
maximum temperature. The proper loop decay time then becomes:
\begin{equation}
\tau_d = \sqrt{\frac{T_0}{T_M}} \tau^\prime_s
\end{equation}
If we use the maximum temperature $T_0$ to estimate the loop length
with expressions derived from Eq.~(\ref{eq:tserio}) (Reale et al. 1997),
instead of $T_M$, we then overestimate the loop length. However, since
the dependence on the temperature is rather weak in Eq.(\ref{eq:tserio})
and the temperatures not very different, we expect not a too large
effect on previous results. The error can be easily estimated from the
square root of the ratio of the maximum temperature to the temperature
at the density maximum. Since this ratio is typically of about 1.2-1.5
(see Sec.\ref{sec:appl}), the amount of overestimate is no more than
15-20\%. Furthermore, if we know this ratio we can correct for this
effect.

\begin{figure}
\centerline{\psfig{figure=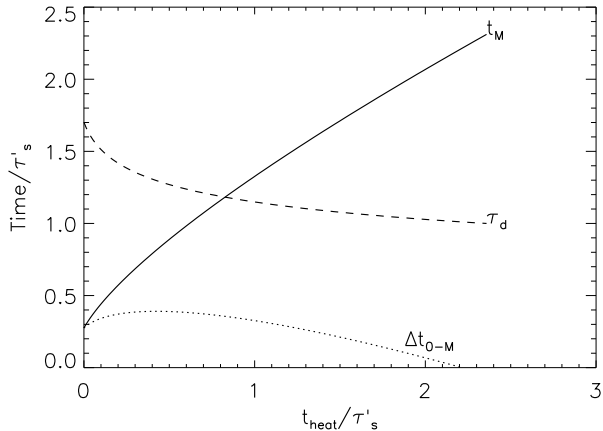,width=9cm}}
\caption[]{Time of the density maximum ($t_{M}$, {\it solid line}), delay
of the beginning of the density decay from that of the temperature decay
($\Delta t_{0-M}$ , {\it dashed line}), and loop decay time ($\tau_{d}$,
{\it dotted line}) vs the duration of the heat pulse $t_{heat}$. All
times are in units of the equilibrium decay time $\tau^\prime_s$.}
\label{fig:times}
\end{figure}

Fig.~\ref{fig:times} shows how the decay time varies with the duration
of the heat pulse. As expected, the decay time is invariably larger
than the equilibrium decay time: the shorter the heat pulse, the longer
is the decay time, with a maximum of about 1.7 $\tau^\prime_s$.  However,
in most of the range, the decay time is larger no more than $20 \%$
than the equilibrium decay time, and therefore the loop length
overestimated by the same amount.

In all the above analysis we have neglected the effect of the star
gravity.  This is reasonable both because the gravity have very little
influence on the strong initial flare dynamics, and because, at the high
temperatures of stellar flares, the pressure scale height:

\[
h_p \sim 5000 \frac{T}{g_*/g_{\odot}} = 50 \frac{T_7}{g_*/g_{\odot}}
\]
is larger than the height of the flaring loops (assuming they stand
vertical on the stellar surface). For very long loops, comparable to
the stellar radius, the pressure scale height is even longer because
the gravity decreases significantly at long distances from the surface,
so that the effect of gravity is expected to be even smaller.

Finally, throughout our analysis we have assumed a simple heating
function: a top-hat in time, uniform distribution in space. Different
heating functions may have some effects on the results (somewhat discussed
in Sec.~\ref{sec:appl}), but except for very extreme cases, most of them
should be smoothed out by the very efficient thermal conduction.

\subsection{Loop length from the rise phase}
\label{sec:len}

Stellar flare observations do not always cover the whole flare duration;
sometimes the rise phase is missing, sometimes most of the decay is not
observed. The latter possibility can occur quite frequently, because the
decay is the longest part of the flare and, for long flares, the
observation may end too early in the decay. One result of the present
work is to provide a set of useful formulae for events with missing
observed decay. This could be obtained ultimately because most of the
rise phase is ruled by the same processes which rule the decay phase,
i.e. the energy losses.
Eq.(\ref{eq:lris}) yields the loop length, from the time of the
maximum density $t_M$ (the emission measure can be used as a proxy),
the maximum temperature $T_0$ and the temperature at the density
maximum $T_M$.  

Alternatively, we may derive the loop length from
Eq.(\ref{eq:ldelt}) even if we miss the flare start but we are
able to measure the time interval between the temperature maximum and
the density (or emission measure) maximum.  If both times are available
to us, we may choose the less uncertain one as the more reliable, and
we may use both of them to check for consistency and to derive a more
accurate length value from a weighted average.
We should in fact consider that the determination of the relevant
flare times can be affected by significant uncertainties.  The time of
the density maximum is typically determined within a time bin where
the spectral fitting is performed. This time bin can be quite large.
Moreover, it is reckoned from the time of the flare start, which can in
turn be not well-determined. Analogously, the uncertainty in the interval
$\Delta t_{0-M}$ comes from the width of both the time bins including
the temperature maximum and the density maximum.  The strong dependence
on the ratio $\psi = T_0/T_M$ can also add to further uncertainties.

Eqs.(\ref{eq:lris}) and (\ref{eq:ldelt}) provide the loop length even if
only the rise and peak phase of the flare are observed.  These expressions
are alternative and independent of the expressions based on the flare
decay and therefore of the presence of any significant heating during
the decay.  They can provide therefore a further check on the loop
length estimation, if both loop length values are available. There is
a chance to obtain inconsistent results from the two approaches if the
flare decay progressively involves other different coronal loops, or
else if the heat pulse triggering the flare is not a top-hat function,
instead, for instance, slowly rising.

If no time-resolved spectral information is available, e.g.  because of
low photon statistics, the time of the light curve maximum may be used
as a proxy of the time of the emission measure maximum. Since the former
occurs a bit earlier than the latter, the loop length estimated from
the former will be slightly underestimated.

\subsection{Loop Cross-section}

A typical output of the analysis of X-ray spectra with moderate
resolution, such as those from CCD detectors (e.g. EPIC/XMM-Newton,
ACIS/Chandra), is the emission measure associated with a fit thermal
component. Time-resolved spectra during the flare can then provide us with
the flare maximum emission measure.  If we know also the maximum density,
we can then derive the loop volume, and, knowing the loop length, also
the loop cross-section area and radius.  From Eq.(\ref{eq:nmax}) and/or
Eq.(\ref{eq:ldelt}) we can evaluate the maximum density at the loop top.
Since the emission measure is integrated over most of the loop (the part
emitting in the instrument band), For consistency, the volume should 
be computed using the density averaged of the emitting part of the loop.
In realistic conditions of pressure equilibrium, this average density
is higher than the
density at the loop apex, because the temperature decreases downwards
from the loop apex.  A reasonable estimate of this average density can
be derived from the expressions linking the loop apex temperature to the
temperature obtained from spectrum fitting (e.g. Favata et al. 2000b),
under the reasonable assumption that the fit temperature is an average
loop temperature:

\begin{equation}
T_M= \xi ~ T_{avg}^{\eta}
\end{equation}
where the parameters $\xi$ and $\eta$ depend on the observation instrument
(e.g. $\xi = 0.130$ and 0.233 and $\eta = 1.16$ and 1.099 for EPIC/XMM-Newton, 
and MECS/BeppoSAX, respectively, see also the Appendix, Eq.~\ref{eq:tmax}). 
Then the average density is:

\begin{equation}
n_{avg} = n_M \frac{T_M}{T_{avg}}
\label{eq:navg}
\end{equation}

Typically $n_{avg}/n_M \sim 1.5 - 2$. The loop volume $V$, cross-section area
$A$ and radius $R$ are then:

\begin{equation}
V \approx \frac{EM}{n_{avg}^2}
\label{eq:vol}
\end{equation}

\begin{equation}
A \approx \frac{V}{2 ~ L}
\label{eq:area}
\end{equation}

\begin{equation}
r \approx \sqrt{\frac{A}{\pi}}
\label{eq:radius}
\end{equation}

The final results of such formulae should be in any way taken with care,
because highly indirect and therefore strongly affected by
the error propagation.


\subsection{Single loop versus multi-loop}

It has been pointed out that large and long stellar flares likely involve
entire coronal regions including multiple loop structures.  How can one
reconcile this remark with the single loop approach followed in this work?
Of course, since telescopes are not powerful enough to resolve the flaring
structures, we cannot give conclusive answers.  However, a few arguments
can support our single loop approach.

If multiple loop structures are involved in the flare, this frequently
occurs in the late phases of a flare. The initial phases of an X-ray flare are
instead quite localized and one can reasonably assume the presence of a
single dominant loop, at least in the rise phase.  This is observed in
solar X-ray flares (e.g. Aschwanden \& Alexander 2001) and supported by the
accurate modeling of a well-observed stellar flare at least in the rise,
peak and early decay (Reale et al. 2004).  The clear evidence of a delay
between the temperature and the density peak in many flares is consistent
with a single loop model.
Even in the later flare
phases, a decay with no significant heating, i.e. with a steep path in
the density-temperature diagram as sometimes observed even in very large
flares (e.g. Favata et al. 2005) suggests strong plasma cooling confined
in single loops.  Arcades and two-ribbon flares are instead characterized
by strong heating which completely dominates the flare decay (e.g. Kopp \&
Poletto 1984) and/or by irregular light curves (Aschwanden \& Alexander
2001, Reale et al. 2004)\footnote{Reale et al. (2004) showed that even
arcades of equal loops can be described as single loops.}.
Recently, multi-thread models have been used to study solar flare evolution
(Warren \& Doschek 2005, Warren 2006). In this case, the hydrodynamics 
of the threads will be described by the results presented here.

\subsection{Sample applications}
\label{sec:appl}

As sample applications of the analysis described above, we have revisited
three stellar flares already studied in the literature: a flare on
Algol observed with BeppoSAX on 30 August 1997 (Schmitt \& Favata 1999,
Favata \& Schmitt 1999), a flare on AB Doradus observed with BeppoSAX on
29 November 1997 (Maggio et al. 2000) and a flare on Proxima Centauri
observed with XMM-Newton on 12 August 2001 (G\"{u}del et al. 2004,
Reale et al. 2004). The first flare is quite long ($\sim 1$ day) and
shows an eclipse in the late decay phase. The first two flares are also
hot (above 100 MK) and quite big, involving an emission measure above
$10^{54}$ cm$^{-3}$.  The last flare is cooler and on a smaller scale,
but observed in great detail, thanks to the short distance of Proxima
Centauri. It has been modelled accurately with detailed time-dependent
hydrodynamic simulations (Reale et al. 2004), obtaining constraints even
on the time and space distribution of the heating.  These flares have
been selected because for all of them the parameters related to the
present analysis are all available with the associated uncertainties,
and the uncertainties themselves are not excessively large.

Table~\ref{tab:appl} shows the results obtained from
Eqs.(\ref{eq:nmax})--(\ref{eq:radius}) for these three flares.  The data
in the first ten rows are derived from the original analysis of the
references cited in the table.  The first four are characteristic times,
the light curve decay time ($\tau_{LC}$), the time of the temperature
maximum ($t_0$), and of the density maximum ($t_M$) reckoned since the
beginning of the flare, and the delay between the temperature and the
density maximum ($\delta t_{0-M}$).  There are then the flare maximum
temperature ($T_{obs}$) and emission measure ($EM$) as derived from
one-temperature fitting, and the ratio of the maximum temperature to
the temperature at the density maximum.  The loop maximum temperature
($T_0$), the half-length ($L$) and the thermodynamic cooling time
($\tau_s$) are derived more indirectly according to empirical formulae
(e.g. Reale et al. 1997, see the Appendix).

The last ten rows show results obtained with the analysis presented in this
work. The first four are densities: the loop equilibrium density ($n_0$)
pertaining to the derived maximum temperature ($T_0$), the actual
maximum density ($n_M$) derived in two different ways, and
the maximum density averaged over the whole loop ($n_{avg}$), derived
from the first maximum density.  Then there are the loop volume ($V$), 
cross-section area ($A$), and radius ($r$).
Finally, we show other three values of the loop length: the first one 
is a refinement of the original length derivation, using the temperature
at the density maximum ($T_M$), instead of the maximum temperature
($T_0$).  Since $T_M \leq T_0$, the new length is invariably smaller than
the original, but not by large factors, as discussed in Sec.\ref{sec:len}
(within 20\%), thus mostly confirming the results of the previous
analyses.  The other two length values are obtained from the analysis
of the rise and peak phase, presented here (Eqs.~[\ref{eq:lris}] and
[\ref{eq:ldelt}]).

The first two flares show a significant delay of the time of the
density maximum from that of the temperature maximum -- an indication
that the heat pulse is relatively short as compared to the loop
characteristic cooling time. Coherently, the ratio of the maximum
temperature to the temperature at the density maximum is relatively large
(e.g. Fig.\ref{fig:mod_evol}).  The flare on Proxima Centauri shows
instead a relatively smaller delay and a coherently smaller temperature
ratio.  The loop length and the decay time are very large for the first
flare ($\sim 10^{12}$ cm and 21 ks, respectively) and much smaller for
the other two.

In all flares, the plasma does not reach equilibrium conditions, and the
maximum density is well below the equilibrium density. Although different,
Eqs.(\ref{eq:nmax}) and (\ref{eq:trc}) yield consistent density values
-- within the (quite large) uncertainties -- for all flares, between a
few $10^{10}$ cm$^{-3}$ and a few $10^{11}$ cm$^{-3}$.  The loop volume
and cross-section parameters are affected by even larger uncertainties;
they are derived from the combination of several quantities and suffer
from the error propagation.  It is nevertheless confirmed the large loop
aspect ratio for the AB Dor flare; the aspect ratio for the other flares
is instead closer to typical solar coronal loops.

The loop length values derived from Eqs.(\ref{eq:lris}) and
(\ref{eq:ldelt}) (reported in the last two rows) are generally consistent
within the uncertainties
with those derived from the analysis of the flare decay. They coherently
yield quite smaller -- although still marginally consistent -- values
for the AB Dor flare.  We may take this as a vague indication that this
flare involved progressively larger structures, coherently with the
evidence of significant heating in the decay.  We may also speculate
that the opposite occurred in the Algol flare, i.e. an initial larger
structure, and later other smaller structures, consistent both with the
significant heating in the decay and with the relatively smaller size
obtained from the analysis of the eclipse.

For the flare on Proxima Centauri, we obtain density values smaller
than those reported in Reale et al. (2004), which are, however, derived
assuming quite a different heating deposition, i.e. concentrated at the
loop footpoints.  The loop aspect derived here is coherently larger than
that reported for loop A in Reale et al. (2004), but if we consider
the errors the results are almost compatible. All the loop lengths
obtained for this flare are consistent with that constrained in Reale
et al. (2004).

We remark that in order to obtain consistent results, it is essential
to take into careful account the uncertainties related to each step
of the analysis.  Table~\ref{tab:appl} shows that the analysis of the
rise and peak phase of stellar flares provides valuable results and can
therefore be usefully and extensively applied.

\begin{table*}
\caption{Analysis of three stellar flaring loops. The parameters in the
upper section are derived from the analysis in the references; those in
the lower section from the equations presented in this work. }
\label{tab:appl}
\begin{center}
\begin{tabular}
{l c c c c c}
\hline
Parameter & Equation$^a$ & Units & Algol/BeppoSAX & AB Dor/BeppoSAX & Prox Cen/XMM \\ 
&&& (Favata \& Schmitt 1999) & (Maggio et al. 2000) & (Reale et al. 2004) \\ 
\hline
$\tau_{LC}$ &(\ref{eq:lreale})& $10^3$ s & $50 \pm 5$ & $3.4 \pm 0.1$ & $4.3 \pm 0.1$ \\
$t_0^b$ && $10^3$ s & $13.5 \pm 4$ & $0.72 \pm 18$ & $0.6 \pm 0.1$ \\
$t_M$ &(\ref{eq:nmax})& $10^3$ s & $34.5 \pm 4$ & $1.12 \pm 0.09$ & $0.8 \pm 0.1$ \\
$\Delta t_{0-M}$ &(\ref{eq:dt0rc})& $10^3$ s & $ 21 \pm 6$ & $0.40 \pm 0.18$ & $0.19 \pm 0.09$ \\
$T_{obs}$ &(\ref{eq:tmax})& $10^7$ K & $14 \pm 2$ & $10 \pm 2$ & $2.6 \pm 0.1$ \\
EM &(\ref{eq:vol})& $10^{54}$ cm$^{-3}$ & 13 & 5 & 0.002 \\
$T_0/T_M$ &(\ref{eq:trc})&& $1.4 \pm 0.2$ & $1.5 \pm 0.3$ & $1.18 \pm 0.07$ \\
$T_0$ &(\ref{eq:tmax})& $10^7$ K & $20 \pm 3$ & $14 \pm 3$ & $4.0 \pm 0.2$ \\
$L^{c}$ &(\ref{eq:lreale})& $10^9$ cm & $800 \pm 300$ & $49 \pm 18 $ & $13 \pm 4$ \\
$\tau_s^d$ &(\ref{eq:tserio})& $10^3$ s & $21 \pm 8$ & $1.5 \pm 0.6$ & $0.8 \pm 0.2$ \\
\hline
$n_0$ &(\ref{eq:n0}) & $10^{10}$ cm$^{-3}$ & $6.5 \pm 3$ & $50 \pm 20$ & $16 \pm 5$ \\
$n_M$& (\ref{eq:nmax}) &  $10^{10}$ cm$^{-3}$ & $3.6 \pm 2$ & $12 \pm 6$ & $5 \pm 2$ \\
$n_M$& (\ref{eq:trc}) &  $10^{10}$ cm$^{-3}$ & $3.1 \pm 1.2$ & $23 \pm 12$ & $7 \pm 2$ \\
$n_{avg}$ & (\ref{eq:navg}) &  $10^{10}$ cm$^{-3}$ & $4.4 \pm 1.8$ & $20 \pm 9$ & $9 \pm 2$ \\
V &(\ref{eq:vol}) & $10^{30}$ cm$^{3}$ & $7000 \pm 5000$ & $330 \pm 20$ & $0.25 \pm 0.1$ \\
A &(\ref{eq:area}) & $10^{20}$ cm$^{2}$ & $44 \pm 30$ & $34 \pm 24$ & $0.10 \pm 0.05$ \\
r &(\ref{eq:radius}) & $10^9$ cm & $37 \pm 13$ & $33 \pm 13$ & $1.8 \pm 0.5$ \\
L$^{e}$ &(\ref{eq:lreale},\ref{eq:trc}) & $10^9$ cm & $670 \pm 250$ & $40 \pm 18$ & $12 \pm 4$ \\
L & (\ref{eq:lris}) & $10^9$ cm & $940 \pm 300$ & $27 \pm 13$ & $7 \pm 5$ \\
L & (\ref{eq:ldelt}) & $10^9$ cm & $1300 \pm 700$ & $21 \pm 14$ & $8 \pm 5$ \\
\hline
\end{tabular}
\end{center}

\noindent
$^a$ - where the parameter is used or evaluated;

\noindent
$^b$ - Time of the temperature maximum (or when the temperature begins to decrease);

\noindent
$^c$ - The maximum temperature is used;

\noindent
$^d$ - Using the original expression in Serio et al. (1991) with $\phi = 1$;

\noindent
$^e$ - The temperature at the density maximum is used;


\end{table*}

\section{Conclusions}
\label{sec:conc}

In this work, we derive useful diagnostic formulae for the analysis of the
rise and peak phase of stellar X-ray flares. The basic starting point is the
realization that a flare is generally triggered by a short-lasting heat
pulse, that the shorter the heat pulse the larger is the delay between the
temperature maximum and the density maximum, and that the characteristic
time scales in the rise phase also scale as the cooling times.  We have
then been able to derive useful expressions for the flaring loop density
and length that can be applied if we can measure the flare rise times
and a few significant temperatures.  These expressions are generally
independent of the analysis of the decay phase.  Therefore, they can be
used to complement and enrich the information coming from the analysis
of the decay phase, i.e. to check for consistency, to obtain constraints
on the loop geometry and even on the evolution of the flare morphology.
Of course, they are particularly useful whenever the analysis of the decay
phase cannot be performed, and we
can equally derive an estimation of the loop length.
Our analysis provides useful diagnostics even when the data in the
rise phase are limited.  The information on the loop aspect represents
a higher level of diagnostics than that available only from the decay
phase, and can therefore improve our knowledge on stellar flares and
related coronal structures.  If a complete set of data is available, the
complete analysis provides redundant information, and the opportunity of
a cross-check;
inconsistent results would
not invalidate the present analysis, rather they would show that the
related events are challenging because they cannot be well-described
with our single-loop/simple-heat-pulse model.  We look forward extensive
application of this analysis to large samples of stellar flares.
The results presented here have application beyond flare loop evolution, such
as the dynamical behavior in active region loops.

\acknowledgements{
The author thanks Paola Testa and Antonio Maggio for useful suggestions.
The author acknowledges support from Agenzia Spaziale Italiana and
Ministero dell'Universit\`a e della Ricerca.}

\begin{appendix}

\section{Review of the loop length from the flare decay}

From the basic work of Serio et al. (1991), and including the
information of the density-temperature diagram (Jakimiec et al. 1992, 
Sylwester et al. 1993) to diagnose residual heating, Reale et al. (1997)
device a general expression of the loop length as a function of the
observed decay time. The formula is essentially an inversion of 
Eq.~(\ref{eq:tserio}) with a factor ($F(\zeta)> 1$) 
which corrects for the presence of the heating:

\begin{equation}
L_9=\frac{\tau_{LC} \sqrt{T_7}}{120 F(\zeta)}  ~~~~~~~~ 
\zeta_{min} < \zeta \leq 
\zeta_{max}
\label{eq:lreale}
\end{equation}
where $\zeta$ is the slope of the decay
in the log(n-T) (or equivalent $\sqrt{EM}-T$) diagram and $\tau_{LC}$ is the
e-folding time of the light curve decay (to 1/10 from the maximum).
and
\begin{equation}
T_7 = \xi ~ \frac{T_{obs}^{\eta}}{10^7}
\label{eq:tmax}
\end{equation}
and $T_{obs}$ the maximum best-fit temperature derived from
spectral fitting of the data.

The correction factor is:
\begin{equation}
F(\zeta) = \frac{c_a}{\zeta-\zeta_a} + q_a 
\end{equation}
where $c_a$, $\zeta_a$, $q_a$ are parameters to be tuned for the instrument
which observes the flare.

Table~\ref{tab:dec} shows the values of the parameters for some
relevant solar and stellar instruments with moderate spectral capabilities.

\begin{table}
\caption{Parameters for the determination of the flare loop length from 
the decay phase.}
\label{tab:dec}
\begin{center}
\begin{tabular}
{l c c c c c c c}
\hline
Instrument & $c_a$ & $\zeta_a$ & $q_a$ & $\zeta_{min}$ & $\zeta_{max}$ &
$\xi$ & $\eta$ \\
\hline
{\small ASCA/SIS} 	& 61   & 0.035 & 0.59 & 0.4  & 1.7 & 0.077 & 1.19 \\
{\small BeppoSAX/MECS}	& 0.68 & 0.3   & 0.7  & 0.4  & 1.7 & 0.233 & 1.099 \\
{\small Chandra/ACIS} 	& 0.63 & 0.32  & 1.41 & 0.32 & 1.5 & 0.068 & 1.20 \\
{\small EXOSAT/ME} 	& 1.3  & 0.4   & 0.8  & 0.4  & 1.8 & 0.195 & 1.117 \\
{\small ROSAT/PSPC}	& 3.67 & 0.3   & 1.61 & 0.3  & 1.8 & 0.173 & 1.163 \\
{\small XMM/EPIC} 	& 0.51 & 0.35  & 1.36 & 0.35 & 1.6 & 0.130 & 1.16 \\
\hline
{\small GOES9} 		& 1.02 & 0.37  & 0.36 & 0.37 & 1.7 & 0.097 & 1.163 \\
{\small Yohkoh/SXT}	& 5.4  & 0.25  & 0.52 & 0.3  & 1.7 & (*)   & (*) \\
\hline
\end{tabular}
\end{center}
(*) Yohkoh/SXT is supposed to resolve the flaring loop and to measure
the temperature at the loop apex.
\end{table}
\end{appendix}

\end{document}